%% file: main.tex
\documentclass[conference]{IEEEtran}
\IEEEoverridecommandlockouts
\usepackage{cite}
\usepackage{amsmath,amssymb,amsfonts}
\usepackage{graphicx}
\usepackage{textcomp}
\usepackage{xcolor}
\usepackage{booktabs} 
\usepackage[bookmarks=false]{hyperref}
\usepackage{hyperref}
\usepackage{threeparttable}
\usepackage{algorithm}
\usepackage{algpseudocode}
\usepackage{enumitem}
\usepackage{multirow}
\usepackage{array}
\usepackage{amsmath}
\usepackage{epsfig}
\usepackage{makecell}
\usepackage{booktabs}
\usepackage{subfigure}

\def\BibTeX{{\rm B\kern-.05em{\sc i\kern-.025em b}\kern-.08em
    T\kern-.1667em\lower.7ex\hbox{E}\kern-.125emX}}

\newcommand{\sysname}{Scope}

\begin{document}

\title{\sysname: A Scalable Merged Pipeline Framework for Multi-Chip-Module NN Accelerators}



\author{%
\small
Zongle Huang$^{1,3}$, Hongyang Jia$^1$, Kaiwei Zou$^{2\text{\dag}}$, Yongpan Liu$^1{^,}^{3\text{\dag}}$\\
$^1$Tsinghua University, $^2$Capital Normal University, $^3$Beijing National Research Center for Information Science and Technology\\
huangzl23@mails.tsinghua.edu.cn, zoukaiwei@cnu.edu.cn, \{hjia, ypliu\}@tsinghua.edu.cn
}

\maketitle

\let\thefootnote\relax
\footnotetext{\dag ~Corresponding Author.}

\input{0-abstract}

\input{1-introduction}

\input{2-background}

\input{3-scheme}

\input{4-space}

\input{5-evaluation}

\input{7-acknowledgement}

\input{8-conclusion}

\bibliographystyle{IEEEtran}
\bibliography{IEEEabrv,ref}

\end{document}

%% file: 0-abstract.tex
\begin{abstract}


Neural network (NN) accelerators with multi-chip-module (MCM) architectures enable integration of massive computation capability; however, they face challenges of computing resource underutilization and off-chip communication overheads. Traditional parallelization schemes for NN inference on MCM architectures, such as intra-layer parallelism and inter-layer pipelining, show incompetency in breaking through both challenges, limiting the scalability of MCM architectures.

We observed that existing works typically deploy layers separately rather than considering them jointly. This underexploited dimension leads to compromises between system computation and communication, thus hindering optimal utilization, especially as hardware/software scale. To address this limitation, we propose Scope, a merged pipeline framework incorporating this overlooked multi-layer dimension, thereby achieving improved throughput and scalability by relaxing tradeoffs between computation, communication and memory costs. This new dimension, however, adds to the complexity of design space exploration (DSE). To tackle this, we develop a series of search algorithms that achieves exponential-to-linear complexity reduction, while identifying solutions that rank in the top 0.05\% of performance. Experiments show that Scope achieves up to 1.73× throughput improvement while maintaining similar energy consumption for ResNet-152 inference compared to state-of-the-art approaches.
\end{abstract}


%% file: 1-introduction.tex
\section{Introduction}
\label{chap:intro}

Neural networks (NNs) have achieved great success in various real-world cognitive applications and grown deeper~\cite{alexnet, VGG, resnet, darknet}. To keep up with ever-growing NN computation and model size, multi-chip-module (MCM) architectures have emerged as a promising approach for scaled-out computation power~\cite{MTPU, Simba, NNBaton, COMB, dojo}. It assembles smaller silicon dice known as {\itshape chiplets} onto a package substrate connected with a network-on-package (NoP). Compared to large monolithic chips \cite{Dadiannao, Tetris, Scaledeep}, MCMs can integrate more transistors by breaking the reticle size limit, while reducing design and fabrication costs by lifting yield. Currently, hundreds to thousands of chiplets can already be integrated together~\cite{dojo, 2048chiplet}. Therefore, they are expected to show remarkable advantages in hardware scalability and cost-effectiveness, especially for large-scale NN-based applications. 

However, when the number of chiplet increases, direct deployment of NN layers on MCMs faces (1) \textit{NoP communication overheads} and, (2) \textit{resource underutilization}, potentially negating the benefits of increased computational capacity. (1) NoP links, designed for longer transmission with wider pitch, exhibit lower bandwidth and energy efficiency than on-chip links in system-on-chips~(SoCs), bringing more severe communication overheads. Research~\cite{Simba} demonstrates that with 32 chiplets, NoP communication latency exceeds computational latency, degrading system performance. (2) Partitioning NN layers with fixed parallelizable dimensions across numerous chiplets results in insufficient workload per chiplet, with typical resource utilization rates below 40\% for 64 chiplets~\cite{Utilization}. These limitations create a substantial gap between MCMs' theoretical peak performance and their actual performance, especially when the system scales.

Some works \cite{DNNBuilder,TGPA} have introduced \textbf{\textit{pipelining}} into MCM scheduling to alleviate these two problems. Each layer is mapped to distinct chiplet \textit{\textbf{regions}} rather than the whole package, thus reducing communication volume and mitigating the underutilization caused by over-partitioning.
However, the discrete nature of chiplet assignment poses challenges for pipeline stage matching in deep NNs. For example, given a network with normalized computational loads of (1,3,8,4), 16 chiplets can be allocated proportionally. Yet with 16 layers, each layer can only be assigned to one chiplet, causing layers with more computation to become pipeline bottlenecks. \textit{Therefore, pipelining for NNs incorporating layers with varied computation is only efficient when NNs are \textbf{shallow}.}

\begin{figure}[b]
  \centering
  \includegraphics[width=\linewidth]{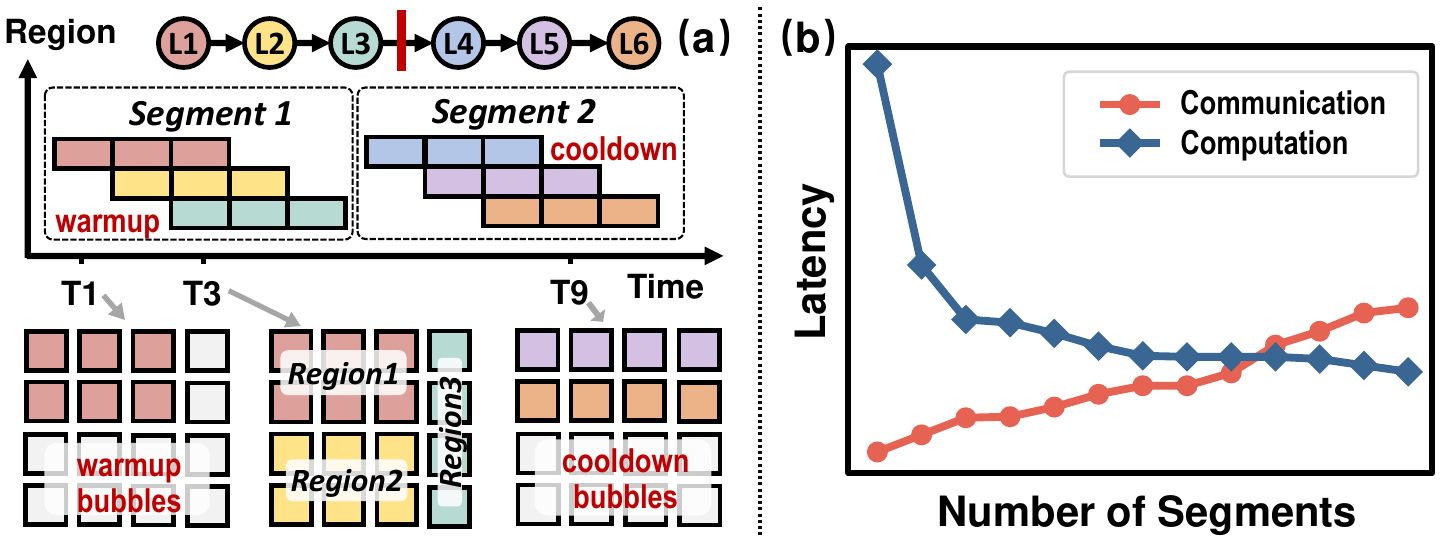}
  \vspace{-8pt}
  \caption{(a) The execution of segmented pipeline. (b) The tradeoff of segmented pipeline. More segments indicate fewer layers are deployed in each sequential deployment, increasing communication overhead, while fewer segments create deeper pipelines, leading to more bubbles and harder stage matching.}
  \label{fig:segmented-pipeline}
\end{figure}

To tackle this, several works \cite{Tangram, DeepBurningSEG, Gemini} further propose \textbf{\textit{segmented pipelining}}, which divides a deep NN into several shallower \textbf{\textit{segments}}. Segments are mapped sequentially to the MCM, while within each segment, layers operate in a pipelined manner across the full package, as shown in Fig.~\ref{fig:segmented-pipeline}(a). By treating each segment as a shallower NN, this approach mitigates the limitations of full pipelining. However, selecting the optimal number of segments presents a trade-off. \textit{Too many segments} indicate that each layer is distributed on many chiplets, incurring the issues of NoP overheads and underutilization; \textit{too few segments} result in each segment containing many layers, leading to a deep pipeline with substantial bubble overhead during the warm-up or cool-down phase, and challenges in achieving stage matching. Fig.~\ref{fig:segmented-pipeline}(b) illustrates this dilemma: more segment counts increase \textit{communication} overheads, while fewer segment counts result in \textit{worse throughput} due to pipeline mismatch and bubbles.

\subsection{Motivation and challenges}


\begin{figure}[t]
 \centering
 \includegraphics[width=\linewidth]{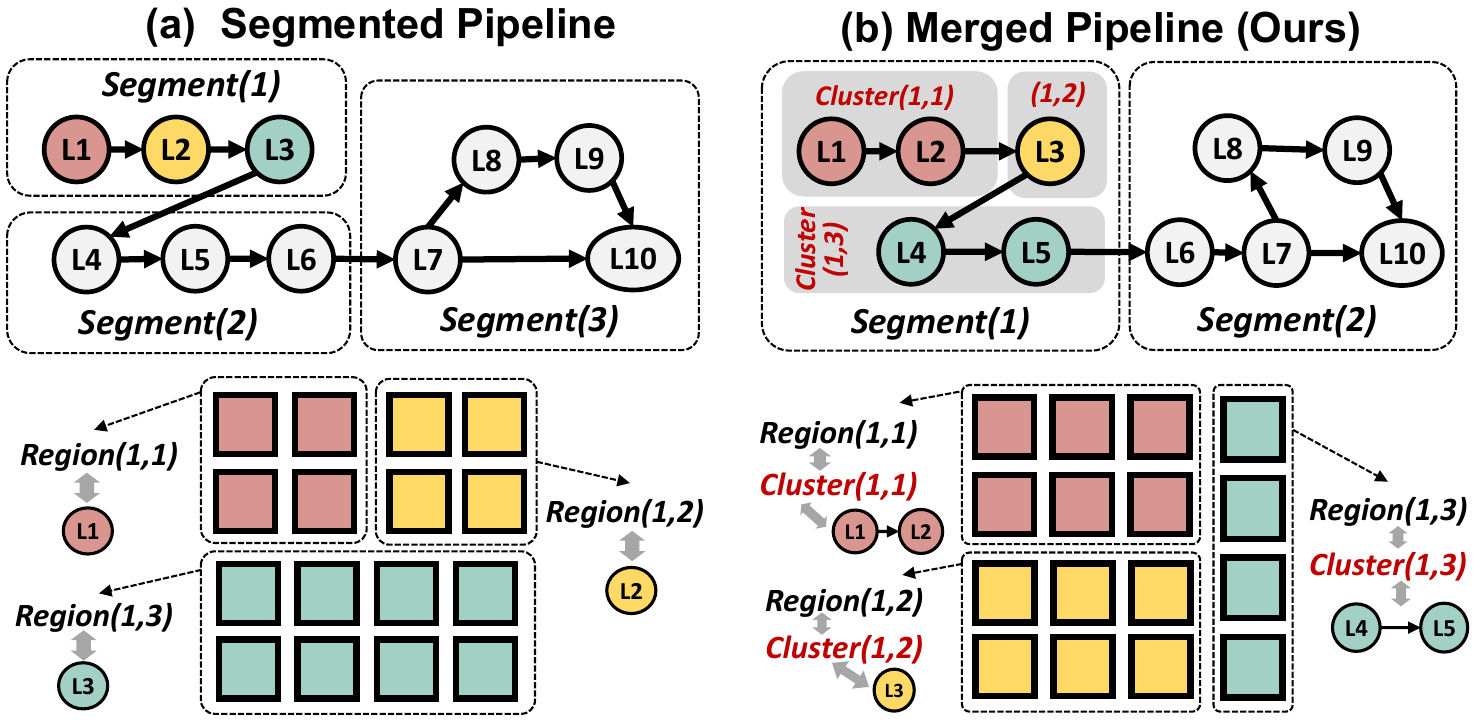}
 \vspace{-8pt}
 \caption{Comparison between segmented pipeline and Scope. We open up the new \textbf{\textit{cluster}} dimension, thus enabling multi-to-multi layer-to-chiplet mapping.}
 \label{fig:compare}
\end{figure}

We found that existing approaches all deploy each layer \textit{\textbf{separately}}. Since each layer has distinct shapes and computational requirements, it is natural that perfectly matching hardware resources to different layers is challenging. This observation motivates our work: by \textit{\textbf{jointly}} considering layers and merging them properly (we call a set of merged layers as a \textit{\textbf{cluster}}), we can create a shallower NN comprising clusters with balanced loads. Rather than \textit{\textbf{distinct}} layers, we then deploy \textit{\textbf{balanced}} clusters on MCM as shown in Fig.~\ref{fig:compare}, which therefore eases hardware allocation and is expected to achieve better utilization. Based on this insight, we develop Scope, a merged pipeline framework that incorporates the cluster dimension into design space exploration (DSE). 

However, the merge pipeline also introduces new challenges. After merging, each segment usually contains more layers, thus requiring saving larger amounts of parameters on the MCM to avoid costly off-chip DRAM access overhead. To address this, we implement distributed storage and on-chip sharing mechanisms to optimize MCM storage efficiency.

The merged pipeline also expands the design search space. While the segmented pipeline maps \textit{\textbf{individual}} layers to \textit{\textbf{multiple}} chiplets, the merged pipeline maps \textit{\textbf{multiple}} layers within a cluster to \textit{\textbf{multiple}} chiplets. The merged pipeline generalizes the segmented pipeline, with the latter being a special case of the former where all clusters only have one layer. To avoid the prohibitive cost of brute force search, we leverage dynamic programming (DP) based on NN layers’ parallelism feature to reduce the algorithmic complexity from exponential to linear. Experimental results demonstrate that this expanded search dimension yields significant performance improvements.

\subsection{Main contributions}
This paper makes the following contributions:

\begin{itemize}[leftmargin=*]
\item We develop Scope, a novel merged pipeline framework for MCM-based NN inference that enhances hardware utilization and reduces NoP overheads, while also mitigating segment count tradeoffs in segmented pipelines.
\item We elaborate flexible layer partitioning and distributed weight storage to further balance computation, inter-chiplet communication and DRAM access during NN execution.
\item We develop a dynamic programming-based search algorithm for DSE that comprehensively considers parallelizable dimensions of NN layers and hardware resource utilization.
\item Experiments show that our search algorithm identifies solutions achieving top 0.05\% performance across the whole search space within an acceptable time. The found schedules exhibit balanced computational loads across clusters and reduce segment counts. Scope achieves 1.73× performance improvement compared to SOTA works.
\end{itemize}


%% file: 2-background.tex
\section{Background}
\label{chap:background}

\subsection{Scalable MCM and Chiplet Architectures}
\label{chap:MCM}

\begin{figure}[t]
  \centering
  \includegraphics[width=0.95\linewidth]{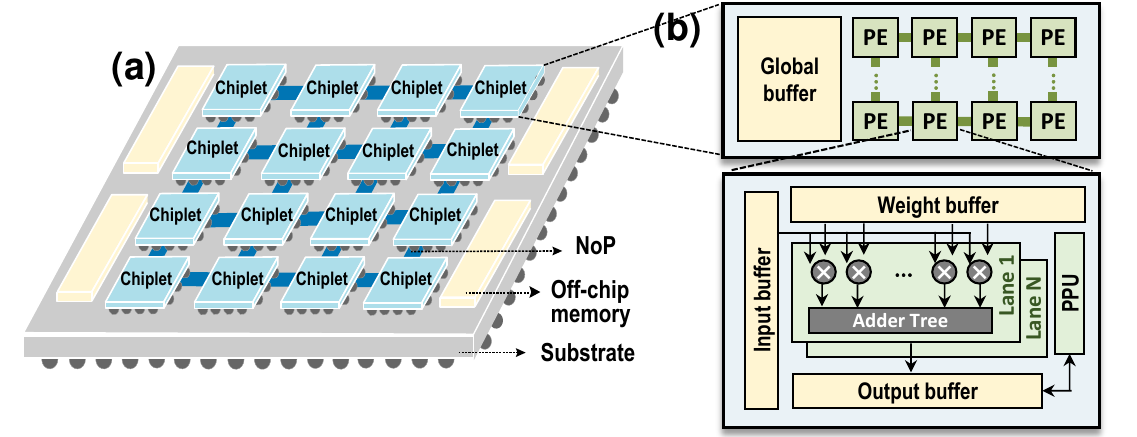}
  \caption{(a) The overview of the MCM structure. (b) The micro-architecture of a chiplet, comtaining hierarchical memory and computing units.}
  \label{fig:architecture}
\end{figure}

MCM achieves scaled-out computational power by integrating multiple chiplets on a silicon interposer \cite{cowos} or an organic substrate \cite{Simba2}. Compared to monolithic multi-tile chips, MCMs offer cost advantages due to higher yield rates of smaller chiplets. MCMs can also integrate more transistors by circumventing the reticle size constraints that limit the area of monolithic silicon chips. However, these benefits come with compromised interconnect performance. Cross-chiplet transmissions require serial/parallel conversion and multi-layer protocol encoding/decoding for reliable transmission, resulting in increased latency, reduced bandwidth, and higher energy-per-bit compared to on-chip interconnects. 

Fig.~\ref{fig:architecture}(a) shows a typical MCM structure. The system adopts the 2D-mesh topology as widely used in previous MCMs~\cite{Simba, dojo, Gemini}. Each chiplet adopts a microarchitecture as shown in Fig.~\ref{fig:architecture}(b), which includes a global buffer~(GB) for activations, a processing element~(PE) array for main computation and parameter buffering, and efficient network-on-chip aggregating results from PEs. PEs are further composed of Lanes performing multiply accumulation~(MAC), adopting the weight stationary dataflow as previous works do~\cite{Simba2}.

\subsection{Intra-layer Partitioning across Chiplets}\label{layerpartition}
\label{chap:partition}

\begin{figure}[t]
  \centering
  \includegraphics[width=\linewidth]{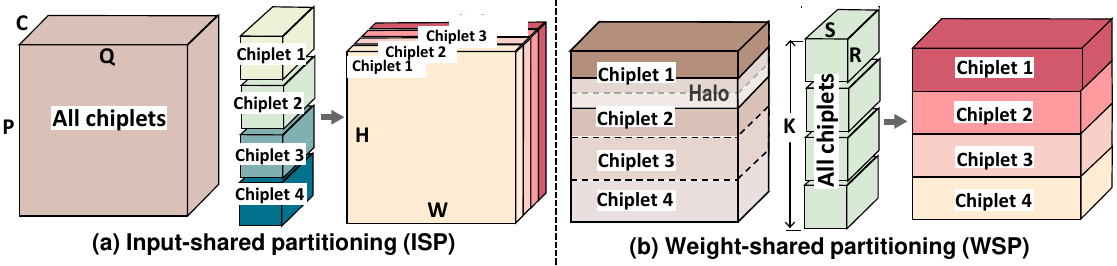}
  \vspace{-5pt}
  \caption{Typical intra-layer partitioning schemes on four chiplets.}
  \label{fig:partition}
\end{figure}

There are three regular workload partitioning approaches to parallel an NN layer: (1)~\textbf{\textit{Input-shared partitioning (ISP)}} as shown in Fig.~\ref{fig:partition}(a), where the inputs are replicated, and the weights are divided proportionally onto the assigned chiplets. ISP introduces NoP communications for activation broadcasting from the previous layer. (2)~\textbf{\textit{Weight-shared partitioning (WSP)}} as shown in Fig.~\ref{fig:partition}(b), where inputs are distributed with each chiplet copying the whole weights. It alleviates the inter-layer communication from the whole activation to only \textit{halo} (the overlapping input regions required due to large kernel sizes), but suffers large runtime weight memory footprint. (3)~\textbf{\textit{Output-shared partitioning (OSP)}}, where both the inputs and weight kernels are partitioned along the input-channel dimension. The OSP partitioning usually incurs higher NoP communications due to the transmission of wide partial sums \cite{Tetris}. Therefore, this work only examines ISP and WSP as previous works usually do \cite{Dadiannao, LearnToScale, NNBaton, Tangram}.

%% file: 3-scheme.tex
\section{The Scope scheme}
\label{chap:scheme}

\subsection{Execution and Modeling}

Fig.~\ref{fig:compare}(b) illustrates the mapping relationships in the merged pipeline, where \textit{\textbf{segments}} and \textit{\textbf{clusters}} are software abstractions, while \textit{\textbf{regions}} are hardware concepts. The system latency $T_{\textit{\text{System}}}$ is the sum of sequentially-performed segment latency $T_{\textit{\text{Segment}}}$ as Equ.~\ref{eq:sys_lat} shows. Tab.~\ref{tab:notation} defines the notation used.
\begin{equation}
    T_{\textit{\text{System}}}=\sum_{i=1}^{N_{\textit{\text{Segment}}}} T_{\textit{\text{Segment(i)}}}
    \label{eq:sys_lat}
\end{equation}

\begin{table}[t]
  \caption{Design parameters and notations}
  \label{tab:notation}
    \centering
    \small
    \setlength{\tabcolsep}{3pt}
    \renewcommand{\arraystretch}{1.3}
    \begin{tabular}{c|l}
    \toprule
    \textbf{Notation} & \textbf{Explanation}\\
   \midrule
    ${\textit{\text{Segment(i)}}}$ &  The $i$th Segment in the whole NN, $i=1,...,N_{\textit{\text{Segment}}}$.\\
    \hline
    ${\textit{\text{Cluster(i,j)}}}$ &  The $j$th Cluster in the $i$th Segment, $j=1,...,N_{\textit{\text{Cluster}}}^i$.\\
    \hline
    ${\textit{\text{Region(i,j)}}}$ &  \multicolumn{1}{m{7cm}}{The Region executing {\textit{\text{Cluster(i,j)}}}. In a Segment, Region counts equal Cluster counts.}\\
    \hline
    ${\textit{\text{Layer(i,j,k)}}}$ & \multicolumn{1}{m{7cm}}{The $k$th Layer in (the $i$th Segment, the $j$th Cluster), $k=1,...,N_{\textit{\text{Layer}}}^{i,j}$.}\\
    \hline
    $P(i,j,k)$ &  \multicolumn{1}{m{7cm}}{The partitioning scheme for the \textit{Layer(i,j,k)}. $P(i,j,k)$ represents ISP or WSP.}\\
  \bottomrule
    \end{tabular}
\end{table}

\begin{figure}[t]
  \centering
  \includegraphics[width=\linewidth]{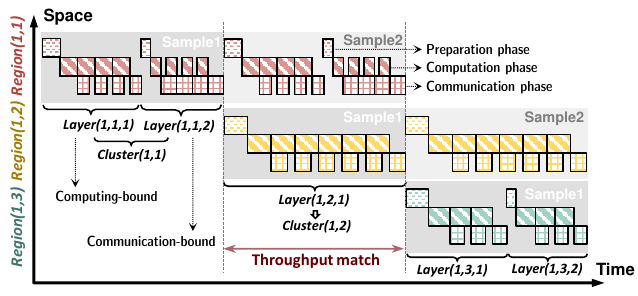}
    \caption{Scope's decomposed execution and timeline within a segment, which is composed of pipelined clusters.  Each cluster can be further broken down into layer-wise execution. We overlap computing phase and communication phase to reduce processing time.}
  \label{fig:pipeline}
\end{figure}

Fig.~\ref{fig:pipeline} illustrates the execution within a single segment, where the processing of different clusters composes distinct pipeline stages. Since $T_{\textit{\text{Segment(i)}}}$ is bottlenecked by the longest stage, given fixed hardware resources, the system achieves optimal throughput when cluster execution times are approximately the same. Given a total of $m$ samples, we can express each Segment's processing time $T_{\textit{\text{Segment(i)}}}$ as:
\begin{equation}
    T_{\textit{\text{Segment(i)}}}=(m + N_{\textit{\text{Cluster}}}^i - 1) \times \mathop{max}\limits_{j=1,...,N_{\textit{\text{Cluster}}}^i}(T_{\textit{\text{Cluster(i,j)}}})
    \label{eq:segment_lat}
\end{equation}

The cluster latency is determined by the execution of its constituent layers, as shown in Equ.~\ref{eq:cluster_lat}. The execution of each NN layer comprises the following three phases.
\begin{equation}
    T_{\textit{\text{Cluster(i,j)}}}= \sum_{k=0}^{N_{\textit{\text{Layer}}}^{i,j}} T_{\textit{\text{Layer(i,j,k)}}}
    \label{eq:cluster_lat}
\end{equation}

\textbf{(1) Preparation Phase.} In this phase, off-chip memory or other chiplets (detailed in Sec.~\ref{chap:shared-buffer}) prepare weight and activation for this layer. Given the MCM platform, its time $T_{\textit{\text{Layer(i,j,k)}}}^{\textit{\text{pre}}}$ is related to the weight volume and how weights are divided among chiplets in the corresponding region, as expressed in Equ.~\ref{eq:fun1}. The function $\mathbb{F_1}$ represents DRAM and NoP behaviors given these tasks, which are regressed from simulators like Ramulator2~\cite{luo2023ramulator2} and BookSim2~\cite{booksim2}.
\begin{equation}
    T_{\textit{\text{Layer(i,j,k)}}}^{\textit{\text{pre}}} = \mathbb{F_1}\big(\ \textit{Layer(i,j,k)},\ \textit{P(i,j,k)},\ \textit{Region(i,j)}\ \big)
    \label{eq:fun1}
\end{equation}

\textbf{(2) Computation Phase.} In this phase, computation is distributedly performed on chiplets following partition methods illustrated in Fig.~\ref{fig:partition}. For both methods, the theoretical operations per chiplet are reduced to $1/||\textit{\text{Region(i,j)}}||$ of the unpartitioned load, where $||\cdot||$ gets the chiplet counts. However, ISP reduces the parallelable weight dimension, potentially impacting resource utilization. Equ.~\ref{eq:fun2} describes above relationships, with $\mathbb{F_2}$ derived from DNN simulator Timeloop~\cite{timeloop}.
\begin{equation}
    T_{\textit{\text{Layer(i,j,k)}}}^{\textit{\text{comp}}} = \mathbb{F_2}\big(\ \textit{Layer(i,j,k)},\ \textit{P(i,j,k)},\ ||\textit{Region(i,j)}||\ \big)
    \label{eq:fun2}
\end{equation}

\textbf{(3) Communication Phase.} In this phase, computation results are collected and distributed to the corresponding regions of next layer, which could be the same region as this layer (Case1, indicating this layer and the next layer belong to the same cluster, like ${\textit{\text{Layer(1,1,1)}}}$ and ${\textit{\text{Layer(1,1,2)}}}$ in Fig.~\ref{fig:pipeline}), or a different region (Case2, indicating they belong to different clusters, like ${\textit{\text{Layer(1,1,2)}}}$ and ${\textit{\text{Layer(1,2,1)}}}$ in Fig.~\ref{fig:pipeline}). The communication volume is affected by the partitioning schemes of both the current and subsequent layers, as listed in Tab.\ref{tab:comm-volume}, where \textit{Halo} represents the volume of overlapped inputs in WSP as shown in Fig.~\ref{fig:partition}, and \textit{Output} denotes output activation volume of this layer. We use Equ.~\ref{eq:fun3} to summarize this phase.
\begin{small}
 \begin{equation}
 \begin{aligned}
T_{\textit{\text{Layer(i,j,k)}}}^{\textit{\text{comm}}} & = \mathbb{F_3}\bigg(\textit{Layer(i,j,k)},\ \textit{P(i,j,k)},\ \textit{Region(i,j)},\ \\ &
\begin{cases}
    \textit{Layer(i,j,k+1)},\ \textit{P(i,j,k+1)},\ \textit{Region(i,j)}\ &,Case1\\
    \textit{Layer(i,j+1,1)},\ \textit{P(i,j+1,1)},\ \textit{Region(i,j+1)}\ &,Case2\\
\end{cases}
\bigg)
\end{aligned}
\label{eq:fun3}
\end{equation}
\end{small}

\begin{table}[t]
\caption{Communication volume between chiplets for varied settings.}
\label{tab:comm-volume}
\renewcommand{\arraystretch}{1.1}
\begin{tabular}{cccc}
\toprule
 & \textbf{This layer} & \textbf{Next layer} & \textbf{Communication volume} \\
\midrule
\multirow{5}{*}[1.5ex]{\textbf{Case1}} & WSP & WSP & \textit{Halo} \\
 & WSP & ISP & $(\|{\textit{\text{Region(i,j)}}}\|-1) \textit{\text{Output}}$ \\
 & ISP & WSP & $(\|{\textit{\text{Region(i,j)}}}\|-1) \textit{\text{Output}}+$\textit{Halo} \\
 & ISP & ISP & $(\|{\textit{\text{Region(i,j)}}}\|-1) \textit{\text{Output}}$ \\
\midrule
\multirow{2}{*}{\textbf{Case2}} & WSP / ISP & WSP & \textit{Output} \\
 & WSP / ISP & ISP & $\|{\textit{\text{Region(i,j+1)}}}\| \textit{\text{Output}}$ \\
\bottomrule
\end{tabular}
\end{table}

To minimize the impact of NoP communication on system performance, we \textit{\textbf{overlap it with computation}}. Rather than waiting for the completion of computing the whole activation, NoP transmission begins as soon as partial results are available from the chiplet, thus creating a two-stage pipeline between computation and NoP communication. Consequently, the layer execution time is bounded by the maximum of either computation or NoP transmission time. Equ.\ref{eq:layer_lat} models this relationship.
\begin{equation}
    T_{\textit{\text{Layer(i,j,k)}}}=T_{\textit{\text{Layer(i,j,k)}}}^{\textit{\text{pre}}} + max(T_{\textit{\text{Layer(i,j,k)}}}^{\textit{\text{comm}}}, T_{\textit{\text{Layer(i,j,k)}}}^{\textit{\text{comp}}})
    \label{eq:layer_lat}
\end{equation}

\subsection{Distributed Weight Buffering}
\label{chap:shared-buffer}

A key challenge in Scope is that it still prefers to buffer weights on-chip; otherwise, DRAM access significantly degrades performance and energy efficiency. This issue usually escalates with WSP storing replicated weights. Scope mitigates this by exploiting distributed weight storage across chiplets. Even for WSP layers, each chiplet stores only a weight tiling instead of the whole weight matrix when it's not their turn to compute. Until the preparation phases of these layers, chiplets exchange their weight tiles to ensure that all chiplets hold a complete data copy. This approach reduces on-chip buffer requirements for clusters containing multiple WSP layers.

%% file: 4-space.tex
\section{Design Space Exploration}

\subsection{Search Space}
Based on the analysis of Sec.~\ref{chap:scheme}, implementing Scope scheme requires determining variables in Table~\ref{tab:notation}. Since how to divide segments has been thoroughly discussed in segmented pipeline scheme, for clarity, the following analysis focuses on how to merge layers into a cluster. Specifically, given a single segment containing $L$ layers and an MCM array with $C$ chiplets, we want to find the best \textit{Cluster}, \textit{Region}, and each layer's \textit{Partition} as listed in Table~\ref{tab:notation}.

If there are $N_{\textit{\text{Cluster}}}$ clusters in a segment, the allocation of clusters and their corresponding regions can be viewed as inserting $N_{\textit{\text{Cluster}}}-1$ division across $L$ layers and $C$ chiplets. This generates $Q$ possible configurations as shown in Equ.~\ref{equ:Q}:
\begin{small}
\begin{equation}
    Q(N_{\textit{\text{Cluster}}};L,C) = { L-1 \choose N_{\textit{\text{Cluster}}}-1 }{ C-1 \choose N_{\textit{\text{Cluster}}}-1 }
    \label{equ:Q}
\end{equation}
\end{small}

Since $N_{\textit{\text{Cluster}}}$ can range from 1 to $L$, the total number of possible cluster-region combinations is $\sum_{i=1}^LQ(i;L,C)$. Considering that each layer can implement either ISP or WSP, then the complete search space introduced by Scope is:
\begin{small}
\begin{equation}
    Q_{\textit{\text{total}}} = 2^{L}\sum_{i=1}^LQ(i;L,C)
\end{equation}
\end{small}

When deploying Resnet-152 on 256 chiplets, which is the largest setting in our experiments, $Q_{\textit{\text{total}}}\approx8.27\times10^{\mathbf{164}}$. Therefore, we need an efficient searching algorithm to find a near-optimal solution for Scope.

\subsection{Searching algorithm}

To address the exponential growth in design space driven by clusters, regions, and partitioning, we propose different search methods for each dimension. The overall algorithm is shown in Alg.~\ref{alg:main}, with explanations for each provided below.

\textit{\textbf{For cluster allocation}}, we leverage the inherent parallelism of NN layers to reduce the search space. Since layers within a cluster are processed by the same region, they should exhibit similar parallelizable dimensions to optimize region utilization. Given $N_{\textit{\text{Cluster}}}$, our goal is to determine the optimal set of \textit{Cluster}. We employ dynamic programming to efficiently achieve this. The algorithm initializes with $N_{\textit{\text{Cluster}}}=L$, where each layer forms its own cluster. We then iteratively merge adjacent clusters with the highest parallelism similarity, documenting each merger in a cluster merge table (CMT). Each iteration decrements $N_{\textit{\text{Cluster}}}$ by 1, and after $L$ iterations, the CMT contains cluster division schemes for all possible values of $N_{\textit{\text{Cluster}}}$. This approach efficiently narrows down the search space by incorporating mapping constraints.

Given cluster allocation, we propose a heuristic method to determine \textbf{\textit{optimal regions}}. Aiming to achieve balanced pipeline stages, the initial step allocates chiplets across regions proportionally to the computational load of their corresponding clusters. Since this initial allocation does not account for communication overhead and utilization issues, it often yields suboptimal results. To improve it, our algorithm iteratively redistributes chiplets from regions with shorter processing times to those with longer processing times until the overall latency cannot be further reduced. After determining the number of chiplets per region, we arrange them on the MCM in a ZigZag pattern—an approach adopted and validated by previous works for its effectiveness~\cite{Tangram}. Experiments show that the optimal region allocation can be found in just a few iterations.

\textbf{\textit{For each layer's partition}}, we leveraged the observation that NN typically exhibits larger activation sizes in shallow layers and larger weight sizes in deep layers. Therefore, to reduce the NoP communication overheads, we apply WSP to earlier layers and ISP to later layers. By reformulating the problem from determining per-layer partitioning to identifying a single WSP-to-ISP transition point, we reduced the computational complexity from exponential to linear.

\begin{algorithm}[t]
\caption{The proposed searching algorithm} 
\label{alg:main} 
\small
\hspace*{0.02in} {\bf Input:} A \textit{Segment} containing $L$ layers, chiplet count $C$;\\
\hspace*{0.02in} {\bf Output:} \textit{Cluster}, \textit{Region}, \textit{Partition}; \textit{minLatency};
\begin{algorithmic}

\State \textit{CMT} = \textbf{GenCMT(}\textit{Segment}\textbf{)}
\State \textit{minLatency} = Infinity
\For{$idx$ from 0 to $L$}
\State \textbf{//} WSP for first \textit{idx} layers, ISP for remaining layers.
\State \textit{tmpPartition[:idx]} = WSP, \textit{tmpPartition[idx:]} = ISP
\For{$N_{\textit{\text{Cluster}}}$ from 1 to ${L}$} 
\State \textit{tmpCluster} = \textit{CMT(}$N_{\textit{\text{Cluster}}}$\textit{)}
\State \textit{tmpRegion} = \textbf{ProportionallyAllocate(}\textit{tmpCluster}\textbf{)}
\State \textit{tmpLatency, tmpClusterLatencyList} = \textbackslash \\
\qquad\qquad\qquad\quad\textbf{Forward(}\textit{tmpPartition, tmpCluster, tmpRegion}\textbf{)}
\State \textbf{//} Heuristically search the best \textit{Region}.
\While {\textit{tmpLatency} $\textless$ \textit{minLatency}}
\State \textit{minLatency} = \textit{tmpLatency}
\State \textit{Cluster} = \textit{tmpCluster}, \textit{Region} = \textit{tmpRegion}
\State \textit{Partition} = \textit{tmpPartition}
\State \textit{maxClusterIdx} = \textbf{max(}\textit{tmpClusterLatencyList}\textbf{)}
\State \textit{minClusterIdx} = \textbf{min(}\textit{tmpClusterLatencyList}\textbf{)}
\State \textit{tmpRegion[maxClusterIdx]} += 1
\State \textit{tmpRegion[minClusterIdx]} -= 1
\State \textit{tmpLatency} = \textbackslash \\
\qquad\qquad\qquad\quad\textbf{Forward(}\textit{tmpPartition}, \textit{tmpCluster}, \textit{tmpRegion}\textbf{)}
\EndWhile
\EndFor
\EndFor
\State return \textit{minLatency, Cluster, Region, Partition}
\end{algorithmic}
~\\
\textbf{GenCMT(}\textit{Segment}\textbf{)}:\quad\textbf{//} Build \textit{cluster merge table~(CMT)}
\begin{algorithmic}
\State \textit{CMT} = \{\}, \textit{CMT[L]} = \textit{Segment}
\For {$N_{\textit{\text{Cluster}}}$ from \textit{L} to 2}
\State\textit{curCluster} = \textit{CMT[$N_{\textit{\text{Cluster}}}$]}
\State\textit{parallel} = \textbf{ComputeParallelism(}\textit{curCluster}\textbf{)}
\State\textit{parallelOffset} = \textbf{abs(}\textit{parallel[:$N_{\textit{\text{Cluster}}}$-1] / parallel[1:]} - 1\textbf{)}
\State\textit{minOffsetIdx} = \textbf{min(}\textit{parallelOffset}\textbf{)}
\State\textit{Cluster} = \textbf{merge(}\textit{curCluster}, \textit{minOffsetIdx}\textbf{)}
\State\textbf{CMT}[$N_{\textit{\text{Cluster}}}$-1] = \textit{Cluster}
\EndFor
\State return \textit{CMT}
\end{algorithmic}
\end{algorithm}

%% file: 5-evaluation.tex
\section{Evaluation}
\label{chap:exp}

\begin{figure}[t]
 \centering
 \includegraphics[width=0.8\linewidth]{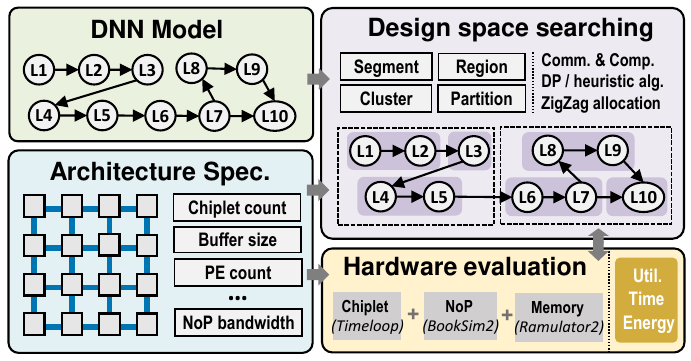}
 \caption{The overview of Scope Framework.}
 \label{fig:overview}
\end{figure}

\begin{figure*}[h]
  \centering
  \includegraphics[width=0.98\textwidth]{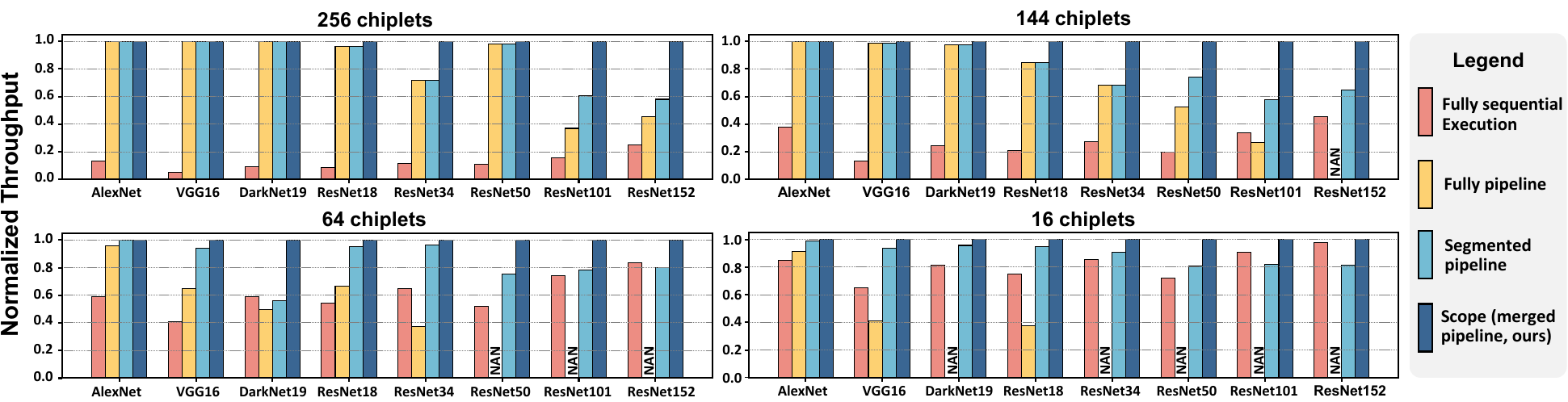}
  \vspace{-10pt}
    \caption{Normalized throughput of different methods when deploying various networks on MCMs of different scales. Scope achieves the most significant performance improvement in the largest-scale experiment (ResNet152 on 256-chiplet MCM).}
  \label{fig:throughput}
\end{figure*}

\subsection{Experiment settings}

The experiments are conducted on a simulated MCM architecture as described in Section \ref{chap:MCM}, with detailed configurations presented in Table \ref{tab:configure}. Energy measurements for digital modules and SRAM are obtained through Synopsys Design Compiler and SRAM Compiler synthesis at 28nm 800MHz. Hardware evaluation for chiplet, NoP, and DRAM leverages results of open-source simulators Timeloop~\cite{timeloop}, BookSim2~\cite{booksim2}, and Ramulator2~\cite{luo2023ramulator2}, respectively. Fig.~\ref{fig:overview} shows the Scope framework, with design variables listed in Table~\ref{tab:notation} obtained through DSE. We evaluate AlexNet, VGG16, DarkNet19, and ResNet18/34/50/101/152, using 8-bit weights/activations and 24-bit accumulation. Our baselines include state-of-the-art works on deploying DNNs on chiplets, whose scheduling can be categorized into three types as introduced in Sec.~\ref{chap:intro}: fully sequential (\cite{Simba,Simba2,NNBaton}), fully pipeline (\cite{DNNBuilder, TGPA}), and segmented pipeline execution (\cite{Tangram, DeepBurningSEG, Gemini}). For a fair comparison, Scope uses an identical segment allocation method as the segmented pipeline to isolate performance gains solely to our novel contributions.

\label{chap:algorithm}

\begin{table}[t]
  \caption{Evaluation Setup}
  \label{tab:configure}
    \centering
    \small
    \setlength{\tabcolsep}{6pt}
    \renewcommand{\arraystretch}{1.1}
    \begin{tabular}{c|l}
    \toprule
    Component & Parameters\\
    \midrule
    Chiplet & \multicolumn{1}{m{6cm}}{4$\times$4 PEs, 8 Lanes per PE, 8 MACs per Lane. 64KB weight buffer per PE, 64 KB global buffer, 0.2pJ for each 8-bit MAC}\\
    \hline
    NoP & \multicolumn{1}{m{6cm}}{2D mesh topology, 100GB/s/chiplet bandwidth, 1.3pJ/bit}\\
    \hline
    Main memory & \multicolumn{1}{m{6cm}}{100GB/s total bandwidth, 128-bit LPDDR5}\\
  \bottomrule
        \hline
    \end{tabular}
\end{table}

\begin{figure}[t]
  \centering
  \includegraphics[width=0.78\linewidth]{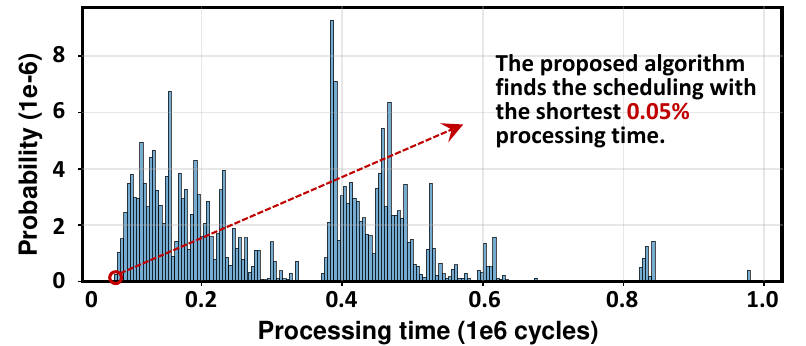}
  \vspace{-8pt}
    \caption{Processing time distribution of all valid scheduling. The x-axis represents the processing time range, while the y-axis shows the proportion of schedules with processing time falling into the corresponding interval.}
    \vspace{-10pt}
  \label{fig:validation}
\end{figure}

\subsection{Experiment Results}


\noindent \textbf{\underline{(1) Validation of Searching Methodology.}}
Since the proposed search algorithm reduces the number of samples requiring simulation across the whole design space, we first validate that high-performance solutions remain attainable within the pruned search space in this subsection. Using AlexNet deployment on a 16-chiplet MCM\footnote{The smallest-scale setting is selected, as experiments with larger scales make exhaustive search computationally prohibitive.}, we compare valid solutions from exhaustive search with our algorithm's output. As illustrated in Fig.~\ref{fig:validation}, our approach identifies the solution with a top 0.05\% performance. The algorithm is also efficient, with searching time of the largest experiment (ResNet152 on 256-chiplet MCM) being approximately one hour on an Intel Core i-13700H processor. The original design space contains $O(10^{164})$ samples, which is impossible to fully search.


\vspace{3pt}
\noindent \textbf{\underline{(2) Performance Comparison. }}
Fig.~\ref{fig:throughput} demonstrates the normalized throughput of different NN tasks at varying MCM scales. Scope consistently achieves optimal performance across all configurations, with maximum gains in the \textbf{deepest} NN utilizing \textbf{most} chiplets, demonstrating up to 1.73x performance improvement compared to SOTA solutions. 

Analysis of baselines are as follows.
(a) Sequential execution exhibits better performance with fewer chiplets, where communication overhead and resource underutilization are minimized. However, as the hardware scales, its performance significantly degrades and becomes the least efficient scheduling. 
(b) Pipeline execution achieves optimal results with increased chip count and shallower networks, facilitating effective pipeline stage matching. However, its performance degrades significantly as network depth increases, even failing to be valid due to weight buffer overflow. 
(c) Segmented pipeline execution demonstrates improvements over the previous two alternatives at scale, establishing it as the existing SOTA approach, outperformed only by Scope. 

\begin{figure}[H]
  \centering
  \includegraphics[width=0.65\linewidth]{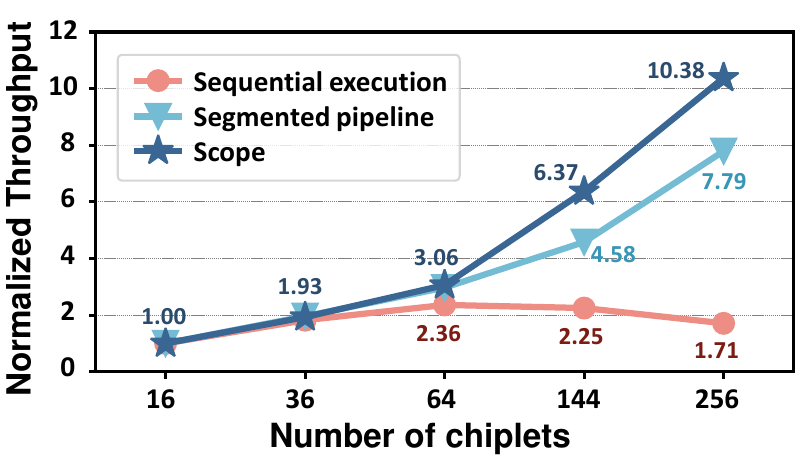}
  \vspace{-8pt}
    \caption{The normalized throughput for different methods as the number of chiplets increases. For each method, the throughput is normalized to the case with 16 chiplets. Scope exhibits the best scalability.}
  \label{fig:scalability}
\end{figure}

\vspace{3pt}
\noindent \textbf{\underline{(c) Scalability Evaluation. }}
Fig.~\ref{fig:scalability} shows the scalability of various methods with increasing chiplet counts under a fixed workload. The fully-pipelined approach is excluded due to a lack of valid solutions at lower chiplet counts. Scope demonstrates the best scalability. The throughput of the segmented pipeline increases more slowly than Scope, while that of the fully-sequential method may even decrease with more chiplets, as NoP communication becomes the bottleneck.

\begin{figure}[H]
  \centering
  \includegraphics[width=0.85\linewidth]{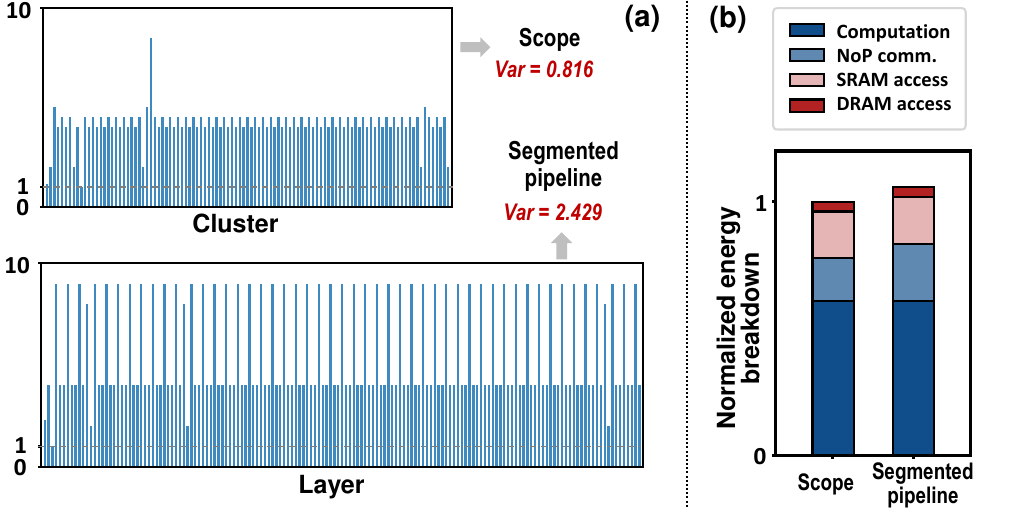}
  \vspace{-8pt}
    \caption{The case study of deploying ResNet152 on a 256-chiplet MCM. a) The normalized computation for clusters in Scope and layers in Segmented pipeline. Scope has a more balanced distribution with smaller variance, thus facilitating easier stage matching. (b) The energy breakdown for both methods, normalized to the total Scope energy consumption.}
  \label{fig:breakdown}
\end{figure}

\vspace{3pt}
\noindent \textbf{\underline{(d) A Case Study of Segmented Pipeline and Scope.}} We conduct a detailed case study comparing segmented pipeline and Scope by analyzing the largest-scale experiment demonstrating the most obvious performance improvement. When processing ResNet152, the segmented pipeline partitions the neural network into \textit{three} segments, while Scope divides it into only \textit{two} segments. Figure~\ref{fig:breakdown}(a) illustrates the computational load across individual layers and clusters. Scope merges layers properly, resulting in balanced workload distribution across clusters, thus facilitating efficient pipeline stage matching. Both methods have roughly equivalent energy consumption and breakdown, as shown in Figure~\ref{fig:breakdown}(b). 
The discrepancy in their latency and energy comparison results from the difference in utilization.

\label{chap:case-study}

%% file: 7-acknowledgement.tex
\section{Acknowledgment}

This work was supported by the National Natural Science Foundation of China (Grant Nos. 92267203 and 62504161) and Deng Feng Fund.

%% file: 8-conclusion.tex
\section{Conclusion}

This paper presents Scope, a coarse-grained pipeline framework introducing a cluster dimension for NN inference orchestration. By merging layers, Scope balances workloads and mitigates pipeline bubbles. Its flexible partitioning and distributed storage optimize computation, communication, and memory access. We devise an efficient linear-complexity algorithm using dynamic programming and heuristics to navigate the design space. Evaluations show Scope achieves up to 1.72× speedup over SOTA and scales better on large MCMs.